\documentclass[twocolumn,showpacs,amsmath,amssymb]{revtex4}
\usepackage{graphicx}
\usepackage{dcolumn}
\usepackage[colorlinks,linkcolor=blue,citecolor=red]{hyperref}
\usepackage{color}
\newcommand{\be}{\begin{equation}}
\newcommand{\ee}{\end{equation}}
\newcommand{\bea}{\begin{eqnarray}}
\newcommand{\eea}{\end{eqnarray}}
\newcommand{\ds}{\displaystyle}

\newcommand{\rr}{{\bf r}}

\begin{document}
\title{Description of longitudinal space charge effects in beams and plasma through dielectric permittivity}

\author{Nikolai Yampolsky}
\email[]{nyampols@lanl.gov}
\affiliation{Los Alamos National Laboratory, Los Alamos, New Mexico, 87545, USA}

\author{Kip Bishofberger}
\affiliation{Los Alamos National Laboratory, Los Alamos, New Mexico, 87545, USA}


\begin{abstract}
We develop a universal framework which allows quickly solve a wide class of problems for longitudinal space charge effects in beams and plasmas in cylindrical geometry. We introduce the longitudinal dielectric permittivity for the beam of charged particles, which describes its collective space charge response. The analyis yields an effective plasma frequency, which depends on the transverse geometry of the system. This dielectric permittivity mirrors the dielectric permittivity of plasma and matches the one dimensional (1D) expression once the transverse size of the beam is large. Several particle species can be included as additive terms describing susceptibility of each specie. The developed approach allows to study stability criteria for collective beam-beam and beam-plasma instabilities for arbitrary transverse distributions in particle densities.
\end{abstract}

\maketitle

\section{Introduction}
\label{sec:Intro}
The problem of collective space charge effects is a major topic of research in plasma and accelerator physics. The research of longitudinal space charge effects covers significant portion of these studies. It ranges from longitudinal space charge waves in plasma columns \cite{trappedplasma} and vacuum tubes \cite{LSCklystron}, beam-cloud instability in ion transport \cite{HeavyIon2Strem}, two-stream instability in vacuum electronics \cite{Carlsten2Stream} and relativistic beams \cite{Jamie2Stream}, longitudinal instability in intense beams \cite{Chao}, microbunching instability in high brightness linacs \cite{microbunching}, free electron lasers in Raman regime \cite{RamanFEL}, plasma wakefield accelerator in capillary \cite{PWFAcapillary} and many others. Despite evident similarity between these problems and similar results obtained by different authors, there is no universal approach to addressing longitudinal space charge effect in various beam/plasma systems in cylindrical geometry. Essentially, every new problem is solved from the first principles repeating similar derivations previously reported in literature.

To date, there are two main approaches to address longitudinal space charge in cylindrical geometry. The first approach is mainly used in  accelerator physics and describes electric fields as a product of impedance and beam current \cite{Chao, JammieZ}.
Another approach originated in plasma physics community \cite{trappedplasma, HeavyIon2Strem, modulation2D, PWFAcapillary}. In that approach longitudinal particle dynamics and Poisson's equation are solved simultaneously, which reduces to problem of finding eigenmodes of the second order differential equation, which represents transverse distribution of electric field in space. A drawback of these approaches is inability to include secondary species of particles such as background plasma or additional beams. Moreover, simple analytic expressions suitable for further studies can be obtained only for limited test transverse distributions, typically flattop \cite{trappedplasma, HeavyIon2Strem, modulation2D, PWFAcapillary} or occasionally Gaussian \cite{PWFAcapillary, SCGaussian}.

Alternatively, collective effects in the medium can be described in terms of polarization density and the resulting electric displacement field. In this approach each specie of particles results in the polarization density independently from other species, reacting to the imposed electric field. As a result, the displacement field of the overall system can be found through additive contribution of each specie. The overall system of multiple beams and plasma species can be described with a single variable, {\it i.e.} dielectric permittivity, in the linear regime when the external electric field is small enough.

\section{Collective longitudinal space charge effects in beams and plasmas}
\label{sec:Analysis}

\begin{figure}[ht]
	\center
	\includegraphics[width=0.45\textwidth]{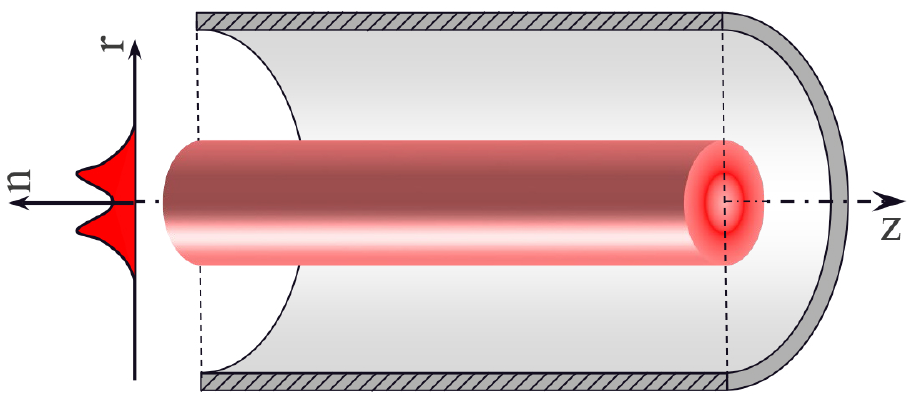}
	\caption{Schematics of the axisymmetric beam propagating inside a vacuum pipe.}
	\label{fig:geometry}
\end{figure}

We consider a long beam of charged particles traveling along $z$ direction inside a beam pipe of circular cross-section as outlined in Fig.~\ref{fig:geometry}. The beam is considered to be axially symmetric and being matched into the focusing channel. The external longitudinal electric field will cause collective response of particles in the beam, which is proportional to the external field in linear regime. The beam density $n(r)$ does not need to be transversely uniform and it is localized close to the pipe axis.

This problem clearly needs to be solved in two dimensions (2D). It is possible to carry out analysis using the formalism developed in Refs.~\cite{HeavyIon2Strem, SC_Geloni}. In that analysis the beam can be viewed as a layered dielectric  with dielectric permittivity matching that of local plasma parameters $\epsilon(r)$. Then the Laplace equation $\nabla (\epsilon(r)\nabla\phi)=0$ for the electrostatic potential $\phi$ can be reduced to a problem of finding an eigen-mode. However, this approach can only be done numerically for complicated enough transverse beam profiles. Moreover, inclusion of additional particle species ({\it e.g.} electron cloud, background plasma, second beam of particles with different energy, {\it etc.}) requires solving different eigen-mode problem.

The external electric field causes density modulation along the beam. That modulation, in turn, results in the induced electric field, which can be viewed as polarization density of the beam.
The induced longitudinal electric field has traverse dependence of its amplitude, which reflects the 3D nature of the problem. However, the transverse scale of the field mirrors that of the longitudinal scale of the beam modulation in the beam frame \cite{GreenFunction}. Therefore, long wavelength modulations in the beam, $k\sigma_r/\gamma \ll1$ will result in transversely uniform longitudinal electric field across the beam. Here $\gamma$ is the relativistic mass factor, $k$ is the longitudinal wavenumber of modulation, and $\sigma_r$ is the rms beam size. Under that condition, the uniform external electric field results in uniform polarization density. As a result,
one can treat the beam as an effective dielectric medium and describe its properties with longitudinal dielectric permittivity $\epsilon_{||}$

\be
\label{epsilon_def}
D_z=\epsilon_{||} E_z.
\ee

The longitudinal dielectric permittivity describes collective space charge effects in the beam.

\subsection{Longitudinal dynamics of particles}

The longitudinal particle dynamics inside the beam can be found using conventional fluid cold plasma equations
\bea
\label{fluid1}
&&\partial_t n +\partial_z (\beta cn)=0,\\
&&(\partial_t+\beta c\partial_z)\left(\frac{\beta}{\sqrt{1-\beta^2}}mc\right)=eE_z,
\label{fluid2}
\eea
where $n$ is the density of charged particles in the beam, $\beta\equiv v_z/c$ is their normalized longitudinal fluid velocity, $m$ and $e$ are the mass and charge of particles, respectively, $\partial_t$ and $\partial_z$ are the partial derivatives over time $t$ and longitudinal coordinate $z$, $c$ is the speed of light. Equations (\ref{fluid1}) --- (\ref{fluid2}) are derived under the assumption that the longitudinal and transverse particle motion are not relativistic in the beam frame, which allows decoupling transverse and longitudinal dynamics. We solve Eqs.~(\ref{fluid1}) --- (\ref{fluid2}) in a linear limit of density and velocity modulations
\be
n=n_0+\delta n,\;\;\;\;\;\;\;\beta=\beta_0+\delta \beta,\;\;\;\;\;\;\left|\frac{\delta n}{n_0}\right|,\left|\frac{\delta \beta}{\beta_0}\right|\ll \frac{1}{\gamma_0^3}.
\ee

We solve linearized fluid equations using Fourier transform
\be 
\label{harmonic}
\delta n, \delta \beta, E\propto e^{i\omega t-ikz}.
\ee 
Then the linearized Eqs.~(\ref{fluid1}) ---(\ref{fluid2}) result in the solution for the perturbation 
\bea
\label{linear1}
&&\delta n=n_0\frac{-ike}{\gamma_0^3m(\omega-k\beta_0c)^2} E_z,\\
&&\delta \beta=\frac{\delta n}{n_0}\frac{\omega-k\beta_0c}{kc}=\frac{-ie}{\gamma_0^3mc(\omega-k\beta_0c)} E_z.
\label{linear2}
\eea

The linearized fluid equations (\ref{linear1}) --- (\ref{linear2}) are correct for arbitrary transverse profiles of the beam density $n_0(r)$ and the applied electric field $E_z(r)$. We consider the case of long wavelength modulations, $k\sigma_r/\gamma_0 \ll1$, which results in a transversely uniform electric field. In this regime the velocity modulation $\delta\beta$ is also transversely uniform (see Eq.~(\ref{linear2})) and the transverse profile of density modulation matches that of the beam profile, $\delta n(r)\propto n_0(r)$ (see Eq.~(\ref{linear1})). This kind of solution ensures that the modulation will not be washed out by transverse particle motion, which mixes particles across the beam. The opposite case of short wavelength modulations, $k\sigma_r/\gamma_0 \gg1$, requires taking  transverse motion into account since each individual particle will observe varying electric field during betatron motion.

\subsection{Space-charge impedance}
The space charge caused by the density modulation described by Eq.~(\ref{linear1}) results in the electric field. Analysis in this section resembles the one presented in Ref.~\cite{CERN_SC} with one major difference. We look for the longitudinal electric field caused by the space charge wave rather than a moving particle. Essentially, the phase velocity of the wave serves as a velocity of the effective particles in the beam. Presence of high density background plasma may significantly reduce velocity of the space charge wave compared to the velocity of charged particles in the beam.  This may cause a major difference compared to conventional analysis in the accelerator physics.

We start with Maxwell's equations for electric and magnetic fields ($E$ and $B$, respectively) in vacuum and include charge and current sources caused by beams ($\rho$ and $j$, respectively)
\bea
\nabla\times{\bf E}&=&-\partial_t{\bf B}/c,\;\;\;\;\;\;\;\;\;\;\;\;\;\;\;\;\;\;\nabla\cdot{\bf E} =4\pi \rho(\rr,t),\\
\nabla\times{\bf B}&=&(4\pi {\bf j}(\rr,t)+\partial_t{\bf E})/c,\;\nabla\cdot{\bf B}=0,\\
{\bf j}&=&e(\delta n {{\bf \beta}_0}c+n_0\delta{\bf \beta}c)\hat{{\bf z}},\;\;\;\;\;\;\,\;\rho=e\delta n.
\eea
These equations can be combined to obtain the second order partial differential equation for the electric field
\be 
\nabla^2{\bf E}-\frac{1}{c^2}\partial^2_{tt}{\bf E}=4\pi\nabla\rho+\frac{4\pi}{c^2}\partial_t{\bf j}.
\ee

We Fourier transform this equation assuming that sources are monochromatic according to Eq.~(\ref{harmonic}). We also take into account that the charge and current densities are related as follows from Eq.~(\ref{fluid1}), $j=\omega/k\cdot\rho$ (it is, essentially, the charge conservation law). Then the equation for the longitudinal component of the electric field becomes
\begin{flalign}
\label{Ez}
\left[\frac{1}{r}\partial_r\left(r\partial_r\right)-k^2+\frac{\omega^2}{c^2}\right]E_z=\frac{4\pi i}{k}\left(k^2-\frac{\omega^2}{c^2}\right)e\delta n.
\end{flalign}

\begin{figure}[ht]
	\center
	\includegraphics[width=0.45\textwidth]{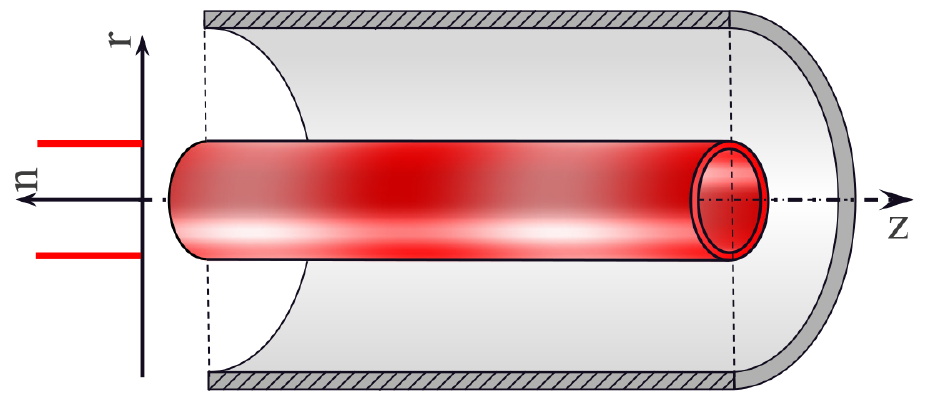}
	\caption{Charge distribution corresponding to Green's function.}
	\label{fig:Green}
\end{figure}

We find the solution of Eq.~(\ref{Ez}) using Green's function approach. We search for the electric field caused by a thin ring of particles with radius $r=r^\prime$, {\it i.e.} $\delta n(r)=\delta(r-r^\prime)$ (Fig.~\ref{fig:Green}). The homogeneous solution of Eq.~(\ref{Ez}) can be represented as linear combination of Bessel functions. These functions are either regular or modified Bessel functions depending on the sign of $k^2-\omega^2/c$. The majority of problems including interaction of beams with plasmas result in a slow space charge wave having phase velocity smaller than the speed of light, $\omega/k<c$ \cite{fast}. In this case the homogeneous solution of Eq.~(\ref{Ez}) can be represented as a linear combination of modified Bessel functions $I_0$ and $K_0$. Then the Green's function $G(r)$ can be found as
\bea
\label{Ez1}
G(r)&=&A_1I_0(\varkappa r)+B_1K_0(\varkappa r),\;\;\;\;\;r<r^\prime,\\
\label{Ez2}
G(r)&=&A_2I_0(\varkappa r)+B_2K_0(\varkappa r),\;\;\;\;\;r>r^\prime,\\
\label{kappa}
\varkappa&=&\sqrt{k^2-\omega^2/c^2}.
\eea

Equations (\ref{Ez1}) --- (\ref{kappa}) should be completed with boundary conditions
\bea
&&|G(r=0)|<\infty,\\
&&\left.G\right|_{r^\prime+\epsilon}=\left.G\right|_{r^\prime-\epsilon},\\
&&\left.\partial_rG\right|_{r^\prime+\epsilon}-\left.\partial_rG\right|_{r^\prime-\epsilon}=4\pi ie\frac{\varkappa^2}{k},\\
&&G(r=a)=0,
\eea
where $a$ is the radius of the vacuum pipe. After some straightforward algebra, we find the amplitude of the electric field on-axis
\begin{flalign}
\left.G\right|_{r=0}=4\pi ier^\prime\frac{\varkappa^2}{k}
\left[\frac{I_0(\varkappa r^\prime)K_0(\varkappa a)}{I_0(\varkappa a)}-K_0(\varkappa r^\prime)\right].
\end{flalign}

The on-axis electric field caused by the space charge of all particles in the beam $E_{SC}$ can be found through the convolution of the particle distribution with the Green's function
\begin{flalign}
\nonumber
&E_{SC}(r=0)=4\pi ie\frac{\varkappa^2}{k}\times\\
&\times\int\limits_0\limits^a\left[\frac{I_0(\varkappa r^\prime)K_0(\varkappa a)}{I_0(\varkappa a)}-K_0(\varkappa r^\prime)\right]\delta n(r^\prime)r^\prime d r^\prime.
\label{solution:Ez}
\end{flalign}

Equation (\ref{solution:Ez}) is similar to the description of space charge in beams using impedance \cite{Chao, JammieZ, CERN_SC, Venturini} since the induced electric field is proportional to particle density, {\it i.e.} current. The similarities and differences are described in details in Sec.~\ref{sec:stationary}.

\subsection{Dielectric permittivity}

The polarization density is proportional to the electric field, $E_{SC}\propto E_z$, as follows from Eq.~(\ref{linear1}) and Eq.~(\ref{solution:Ez}). That allows to introduce the effective dielectric permittivity for the medium
\begin{flalign}
\label{epsilon}
\epsilon_{||}&=\frac{E_z+E_{SC}}{E_z}
=1-\frac{\tilde{\omega}_p^2}{(\omega-k\beta_0c)^2},\\
\tilde{\omega}_p^2&=\varkappa^2\int\limits_0\limits^a\left[K_0(\varkappa r)-\frac{I_0(\varkappa r)K_0(\varkappa a)}{I_0(\varkappa a)}\right] \omega_p^2(r) rdr,
\label{omegapeff}
\end{flalign}
where $\omega_p(r)=\sqrt{4\pi e^2n_0(r)/(\gamma_0^3m)}$ is the relativistic plasma frequency of the beam.

Equation (\ref{epsilon}) shows that the dielectric permittivity of the beam has the same functional dependence as of a uniform plasma. At the same time, the effective plasma frequency $\tilde{\omega}_p$ depends on the geometry of the problem: beam pipe radius, beam density distribution and the effective transverse wavelength of modulation $\varkappa$. Note that  presence of a pipe always results in the reduced value for the effective plasma frequency since the second term in brackets in Eq.~(\ref{omegapeff}) (the term which depends on the pipe radius $a$) is always negative. The effective plasma frequency can be significantly smaller than the characteristic on-axis plasma frequency $\omega_p(0)$ in case of a long wavelength modulation. That reflects the fact that the space charge field is mostly transverse in this case and longitudinal particle interaction is strongly reduced compared to 1D case. The effective plasma frequency can be complex if the resulting system is unstable ({\it e.g.} self-modulation or two-stream instability).

\subsection{Applicability limits}
As discussed above, the results for the description of collective space charge effects using dielectric permittivity (\ref{epsilon}) are strictly valid only when the the longitudinal electric field is transversely uniform across the beam. That is achieved in the limit $\varkappa r_b\ll1$. In general case, extending findings beyond that limit is not appropriate. For example, consider a short wavelength longitudinal space charge wave excited in such a beam. To  first order, the excited space charge wave follows the local dispersion relation $\epsilon_{||}(r)=0$.  The frequency of the longitudinal wave parametrically depends on radius. That causes dephasing between waves at different transverse locations. Transverse mixing of particles causes destructive interference of the waves resulting in Landau damping \cite{Chao}. Moreover, longitudinal and transverse dynamics are parametrically coupled, which can result in the parametric instability \cite{parametric}.

However, the application of longitudinal permittivity of the beam can be cautiously extended into the regime, where $\varkappa r_b\gtrsim1$. Equation (\ref{epsilon}) can be used as an estimate for the permittivity on axis. This formalism can be used if transverse mixing of particles in the beam is limited. For example, it can be achieved in a laminar flow or when the beta function of the beam is the largest scale for the beam dynamics ({\it e.g.} it is larger than the growth rate of the resulting instability). In addition, plasma or beams can be strongly magnetized so that the gyroradius of particles is much smaller than their transverse sizes.

Not that solution for the Maxwell equations described with Eq.~(\ref{solution:Ez}) is obtained without any assumptions for electrostatic limits for the space charge fields. It is valid at any frequencies and wavenumbers, even for superluminal waves having phase velocity larger than the speed of light. In this case the transverse wavevector $\varkappa$ is imaginary and the modified Bessel functions $I_0$ and $K_0$ of imaginary arguments can be rewritten as regular Bessel functions $J_0$ and $Y_0$ of real arguments.

\section{Limiting cases}
\label{sec:cases}
In this section we present several important limiting cases which often serve as baseline models in various studies. 

The effective plasma frequency can be found explicitly for some beam profiles. The results are presented in Table~\ref{table:free}. In that table $\omega_{p0}\equiv\omega_p(r=0)$ is the on-axis plasma frequency, $I=\int en(r)\beta_0 cd^2r$ is full current of the beam, $I_a=mc^3/e\approx 17\,kA$ is Alfven current, $\Gamma(0,x)$ is the upper incomplete gamma function, and  $\gamma_E\approx0.5772$ is Euler–Mascheroni constant. The results for the constant density distribution \cite{VacuumOmegap,microbunching, Venturini} and Gaussian beam profile \cite{SCGaussian, Venturini} reproduce the results reported in the literature.

\begin{widetext}
\onecolumngrid
\begin{table}[h]
\begin{ruledtabular}
\caption{Effective plasma frequency for beams with different density distributions}
\label{table:free}
\begin{tabular}{llll}
 {\bf Density distribution}& 
 ${\bf \tilde{\omega}_p^2/\omega_{p0}^2},\;\;\;\forall\,\varkappa a,\varkappa r_b$&  
$\ds {\bf\tilde{\omega}_p^2/\left(\frac{I}{I_a}\frac{\varkappa^2c^2}{\beta_0 \gamma_0^3}\right)},\;\;\;1/a\ll\varkappa \ll1/r_b$\\ 
\hline
$\ds n(r)=n_0,\;\;\;r\leq r_b$  & 
$\ds 1-{\varkappa r_b}\left(K_1\left({\varkappa r_b}\right)+\frac{I_1(\varkappa r_b)K_0(\varkappa a)}{I_0(\varkappa a)}\right)$&
$-2\ln\left({\varkappa r_b}/2\right)+1-2\gamma_E$\\
$\ds n(r)=n_0\left(1-\frac{r^2}{r_b^2}\right)$&
$\ds 1-4\left({\varkappa  r_b}\right)^{-2}+2K_2\left({\varkappa r_b}\right)-\frac{2I_2(\varkappa r_b)K_0(\varkappa a)}{I_0(\varkappa a)}$&
$-2\ln\left({\varkappa r_b}/2\right)+1.5-2\gamma_E$\\
$\ds n(r)=n_0\exp\left(-\frac{r^2}{2\sigma_r^2}\right)$  &
$\ds \frac{\varkappa^2\sigma_b^2}{2}\exp\left(\frac{\varkappa^2\sigma_b^2}{2}\right)\left[\Gamma\left(0,\frac{\varkappa^2\sigma_b^2}{2}\right)-2\frac{K_0(\varkappa a)}{I_0(\varkappa a)}\right]$&
$-2\ln\left({\varkappa\sigma_r}/2\right)-\gamma_E$\\
\end{tabular}
\end{ruledtabular}
\end{table}
\twocolumngrid
\end{widetext}

\subsection{Quasi-stationary waves in beam frame}
\label{sec:stationary}
Equation (\ref{solution:Ez}) essentially describes on-axis space charge impedance of the beam, $E_z(\omega,k)=Z(\omega,k)I(\omega,k)$. However, the result seem to be different from what is reported in the literature \cite{CERN_SC, JammieZ, Venturini}. The difference comes from source for the electric field being a space charge wave rather than a moving particle. Moving particles can be viewed as a space charge wave, which is stationary in the beam frame. That wave has the dispersion relation in the lab frame $\omega=\beta c k$. As a result, the transverse wavenumber defined in Eq.~(\ref{kappa}) is equal to $\varkappa=k/\gamma_0$ and conventional results for the space charge impedance in beams are recovered.

This approximation can be used if relevant dynamics can be approximated as a quasi-static process in the beam frame ({\it e.g.} \cite{microbunching,Carlsten2Stream,Jamie2Stream,RamanFEL,Tajima}). The exact condition for this approximation is for the phase velocity of the wave in the beam frame to be much slower than the speed of light. The phase velocity of the wave in the beam frame can be found through Lorentz transform of the wave 4-vector
\be
\varkappa\approx\frac{k}{\gamma},\;\;\;\;\;\;\;\;\;\;\;
\left|\frac{\omega^\prime}{k^\prime c}\right|=\left|\frac{\omega-\beta k c}{kc-\beta\omega}\right|\ll1,
\ee
where $\omega^\prime$ and $k^\prime$ are the wave frequency and wavenumber in the beam frame, respectively.

\subsection{Beam in free space}
In this limit the pipe is considered to be so wide that its effect on the beam is not relevant. This limit can be achieved for small enough wavelengths of the modulation, $\varkappa a\gg 1$. This condition allows us to simplify Eq.~(\ref{omegapeff}) for the effective plasma frequency of the beam since the effect of the vacuum pipe becomes negligible, $K_0(\varkappa a)/I_0(\varkappa a)\ll1$
\be
\label{free1}
\tilde{\omega}_p^2=\varkappa^2 \int\limits_0\limits^\infty \omega_p^2(r)K_0\left({\varkappa r}\right)r\,dr,
\ee

 The beam susceptibility in the long wavelength limit $\varkappa r_b\ll1$ is proportional to the overall beam current (last column in Table~\ref{table:free}). Beams with identical currents but different distribution of current in the cross section have similar scaling of the effective plasma frequency, $\tilde{\omega}_p^2\propto \varkappa^2$. At the same time, the numerical pre-factor in the scaling depends on the beams profile, and may vary for moderate ratios of wavelength and beam size, $\ln(\varkappa r_b)\sim 1$.
 
The beam susceptibility in the short wavelength limit $\varkappa r_b\gg1$ recovers 1D limit of uniform plasma since
\be
\tilde{\omega}_p^2\approx\omega_{p0}^2\varkappa^2\int\limits_0\limits^\infty K_0(\varkappa r)rdr=\omega_{p0}^2.
\ee
 
\begin{figure}[ht]
	\center
	\includegraphics[width=0.45\textwidth]{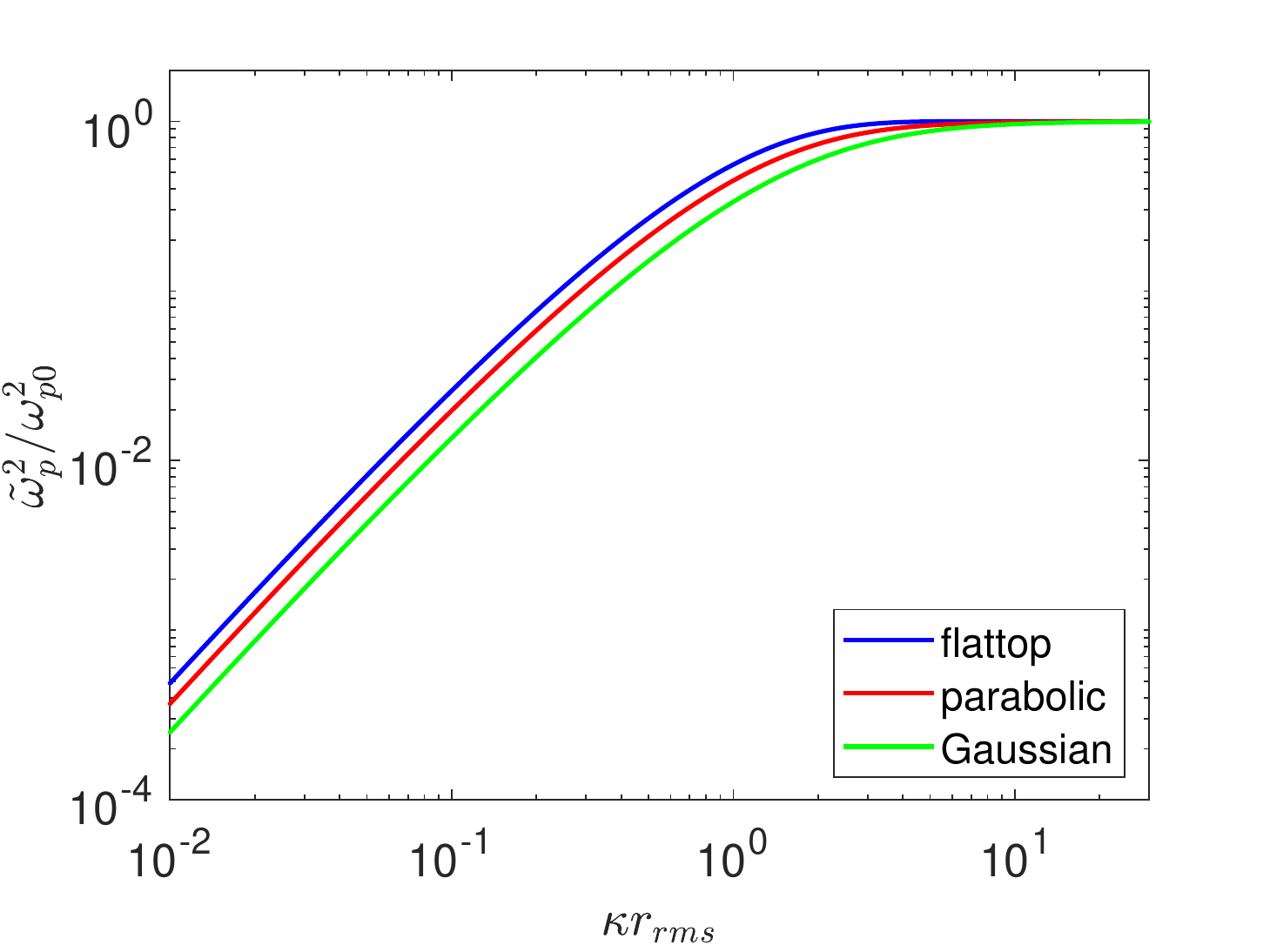}
	\caption{The effective plasma frequency for the flattop (blue), parabolic (red), and Gaussian (green) beam profiles with identical on-axis density and rms radii.}
	\label{fig:free}
\end{figure}

The effective plasma frequency for different beam profiles is presented in Fig.~\ref{fig:free}. The effective plasma frequency matches the 1D solution at small wavelengths, $\tilde{\omega}_p\approx\omega_p$ at $\varkappa r_{rms}\gg1$. The effective plasma frequency reduces while the wavelength of modulation increases. It is a universal scaling that follows from the 2D geometry of the problem. The distributions presented in Fig.~\ref{fig:free} have identical on-axis density and identical root mean square (rms) radii, $r_{rms}^2=\int r^2n(r) rdr/\int n(r) rdr$. Beams with different profiles have the same effective plasma frequency in the 1D limit $\varkappa r_{rms}\gg1$ but have somewhat different effective plasma frequencies at long wavelengths, $\varkappa r_{rms}\ll1$, since their total currents are different from each other.

\subsection{Long wavelength limit}

In this limit the effective transverse wavelength of the modulation is much larger compared to the pipe radius, $\varkappa a\ll 1$. The size of the beam is smaller that the pipe radius, so $\varkappa r\ll1$ as well. The details in the transverse beam distribution are not important in this regime and one can approximate the beam with any test distribution. We choose the beam to have a flattop distribution of density for the simplicity of following derivations.

Then the integrals in Eq.~(\ref{omegapeff}) can be evaluated using asymptotic expansions for the functions in the kernel and we find
\bea
\nonumber
\tilde{\omega}^2_p&\approx&\int\limits_0\limits^{r_b} \left[K_0(\varkappa r)-K_0(\varkappa a)\right]\omega_{p0}^2 r dr\approx\\
&\approx&\frac{\varkappa^2r_b^2}{2}\left[\ln\left(\frac{a}{r_b}\right)+\frac{1}{2}\right]\omega_p^2(0).
\label{longfinal}
\eea
The result matches precisely the one reported in the literature \cite{Chao}.

\section{Generalization of results}
\subsection{Kinetic effects}
The model can be easily expanded to include finite longitudinal energy spread in the beam. This can be done using the kinetic equation for particle distribution in phase space, $f(r,v_z)=n(r)f_0(v_z)$, same way it is typically done for homogeneous plasma. The fluid equations of motion (\ref{fluid1}) --- (\ref{fluid2}) but do not affect the expression for the resulting electric field Eq.~(\ref{solution:Ez}) which can be viewed as a modification to the Laplace equation to include 2D geometry of the beam. Then the longitudinal dielectric permittivity of the beam including kinetic effects  can be described as
\be
\epsilon_{||}=1+\frac{\tilde{\omega}_p^2}{k}\int\limits_L\frac{\partial_{v_z} f_0(v_z)}{\omega-kv_z}dv_z,
\ee
where $f_0(v)$ is particle distribution in velocity, and $L$ indicates integration over Landau contour. This expression is valid for quasi-mono-energetic beams which only have non-relativistic particles in the beam frame, $\gamma_0^3 |\beta_0\delta\Delta\beta|\ll1$.

\subsection{Multiple species of particles}
The resulting expression for the beam permittivity can be easily modified to include other beams or the background plasma. 
Multiple species of particles ({\it e.g.} beams of particles with different masses, charges, or average velocities) can be included as additive terms to the dielectric permittivity. Dielectric susceptibility for each specie describes the polarization of that beam in response to the external electric field $E_z$. The external electric field is identical for each specie, so its polarization  does not depend on whether other types of particles are present. At the same time, the dielectric displacement $D_z$ includes polarization from each type of particles present in the system
\be 
\epsilon_{||}=1-\sum\limits_\alpha \frac{\tilde{\omega}_{p\alpha}^2}{(\omega-k\beta_\alpha c)^2},
\ee
where summation carries over all species $\alpha$ present in the system.

Note that the effective plasma frequency for each specie $\tilde{\omega}_{p\alpha}$ should be calculated using the same transverse wavenumber $\varkappa=\sqrt{k^2-\omega^2/c^2}$ regardless of the energy of each specie. As a result, different species of particles may fall into different limiting cases described in Sec.~\ref{sec:cases}. For example, analysis of beam-cloud instability for the beam propagating through background plasma may result in dielectric susceptibility for the beam being well approximated in the ``free space long wavelength'' limit ($\varkappa r_b\ll1$) while dielectric susceptibility for the background plasma being well approximated in the ``1D'' limit ($\varkappa a\gg1$).

\section{Sample problem: two stream instability}

Several groups have proposed to use two-stream instability for beam bunching in vacuum electronics \cite{Carlsten2Stream, Iran2Stream} or relativistic electron beams \cite{Jamie2Stream,SingleBunch}. That scheme requires presence of electrons with two or more distinct energies co-propagating inside a focusing channel. The kinetic instability similar to the conventional two-stream instability in plasma develops, which results in beam bunching. The analysis of those schemes is typically done in 1D geometry under assumption that the wavelength of modulation in the beam frame is much smaller than the transverse beam size. However, this approximation is not valid in most relativistic cases \cite{Jamie2Stream,SingleBunch}. The transverse beam profile also needs to be accounted for if more complicated geometry of beams is used \cite{{AnnularGun}}.

We consider two beams with close velocities $\beta_1$ and $\beta_2$ co-propagating along a focusing channel. Velocities of these beams are close to each other, so that
\bea
&&\bar{\beta}=(\beta_1+\beta_2)/2,\\
&&\Delta=c(\beta_1-\beta_2)/2=\frac{E_1-E_2}{mc}\frac{1}{2\bar{\beta}\gamma^3}\ll c\bar{\beta}.
\eea
We consider two beams to have identical transverse density profiles and currents for simplicity. The 1D analysis of the instability \cite{Carlsten2Stream,Jamie2Stream} suggests rough scaling for the growth rate of the instability $Im(\omega^\prime)=\gamma Im(\omega-k\beta c)\sim\omega_p$ and the wavenumber for the fastest growing mode $k^\prime=\gamma(k-\beta\omega/c)\sim\omega_p/\gamma^2 (E_1+E_2)/(E_1-E_2)$. As a result, the unstable wave can be viewed as quasi-stationary in the beam frame as discussed in Sec.~\ref{sec:stationary}. The effective plasma frequency depends only on wavelength of modulation since $\varkappa\approx k/\gamma$ in this regime.

The dielectric permittivity for the system of two beams can be found as the additive contribution of two individual beams
\be 
\epsilon_{||}=1-\frac{\tilde{\omega}_p^2}{(\omega-k\bar{\beta}c-k\Delta)^2}-\frac{\tilde{\omega}_p^2}{(\omega-k\bar{\beta}c+k\Delta)^2}.
\ee
We search for the electrostatic plasma waves, which can be supported by this dielectric medium. The electrostatic modes satisfy the dispertion relation $\epsilon_{||}=0$. We introduce the frequency $\Omega=\omega-k\bar{\beta} c$, which describes time evolution in a frame co-moving with average beam velocity. Then this frequency can be found to be
\be 
\label{Omega}
\Omega^2=\tilde{\omega}_p^2+k^2\Delta^2\pm\sqrt{\tilde{\omega}_p^4+4\tilde{\omega}_p^2k^2\Delta^2}.
\ee

In the dispersion relation described with Eq.~(\ref{Omega}), the branch with the minus sign, corresponds to the unstable mode. The instability occurs at large enough wavelengths, $k^2\Delta^2<2\tilde{\omega}_p^2$ . However, the effective plasma frequency scales with the wavenumber. In fact, the vacuum pipe strongly suppresses large wavelength modes. We use the expression (\ref{longfinal}) for the effective plasma frequency to find the condition for the two stream instability to develop
\be 
\label{unstablecondition}
\frac{E_1-E_2}{mc^2}<\sqrt{16\beta\gamma\frac{I}{I_a} \left[\ln\left(\frac{a}{r_b}\right)+\frac{1}{2}\right] }.
\ee
If the beams are not intense enough or their energies are not sufficiently close to each other, then the two-stream instability does not develop. The maximum growth rate for the two-stream instability can be found ($\partial_{k^2}\Omega^2=0$) if the beams are intense enough:
\bea
\label{kmax}
&&\left[\sqrt{1+4\frac{k^2\Delta^2}{\tilde{\omega}_p^2}}-1-2\frac{k^2\Delta^2}{\tilde{\omega}_p^2}\right]\frac{\partial_{k^2}\tilde{\omega}_p^2}{\Delta^2}=\\
&&=2-\sqrt{1+4\frac{k^2\Delta^2}{\tilde{\omega}_p^2}}.
\eea

In the 1D limit the condition for the fastest growing mode reduces to $k^2\Delta^2=3/4\tilde{\omega}_{p0}^2$ since the effective plasma frequency does not depend on the wavelength as illustrated in Fig.~\ref{fig:free}. This result matches findings of other studies \cite{Jamie2Stream}. 

The 2D effects need to be accounted for when the wavelength of modulation in the beam frame is comparable or larger than the beam radius, $kr_b/\gamma\sim1$. At the same time, the effect of the pipe wall is small far enough from the threshold condition (\ref{unstablecondition}) and the beam can be approximated as propagating in free space. We approximate he beams to have flattop distribution of density and use expression for the plasma wave presented in Table~\ref{table:free}. Then the fastest growing mode and the growth rate of instability $Im(\Omega)$ can be found as solutions of the following transcendental equation
\begin{flalign}
\nonumber
&\frac{p^2}{2}K_0(p\kappa)\left[\sqrt{1+\frac{4\kappa^2}{1-p\kappa K_1(p\kappa)}}-1-2\kappa^2\right]=\\
&\;\;\;\;\;\;\;2-\sqrt{1+\frac{4\kappa^2}{1-p\kappa K_1(p\kappa)}},\\
\nonumber
&\frac{\Omega^2}{\omega_{p0}^2}=1-p\kappa K_1(p\kappa)+\kappa^2-\\
&\;\;\;\;\;\;\;-\sqrt{(1-p\kappa K_1(p\kappa))^2+4\kappa^2(1-p\kappa K_1(p\kappa)},
\end{flalign}
where
\be
\kappa=\frac{k\Delta}{\omega_{p0}},\;\;\;\;\;\;\;p=\frac{r_b\omega_{p0}}{\gamma\Delta}=\frac{mc^2}{E_1-E_2}\sqrt{16\beta\gamma\frac{I}{I_a}}.
\ee

\begin{figure}[ht]
	\center
	\includegraphics[width=0.450\textwidth]{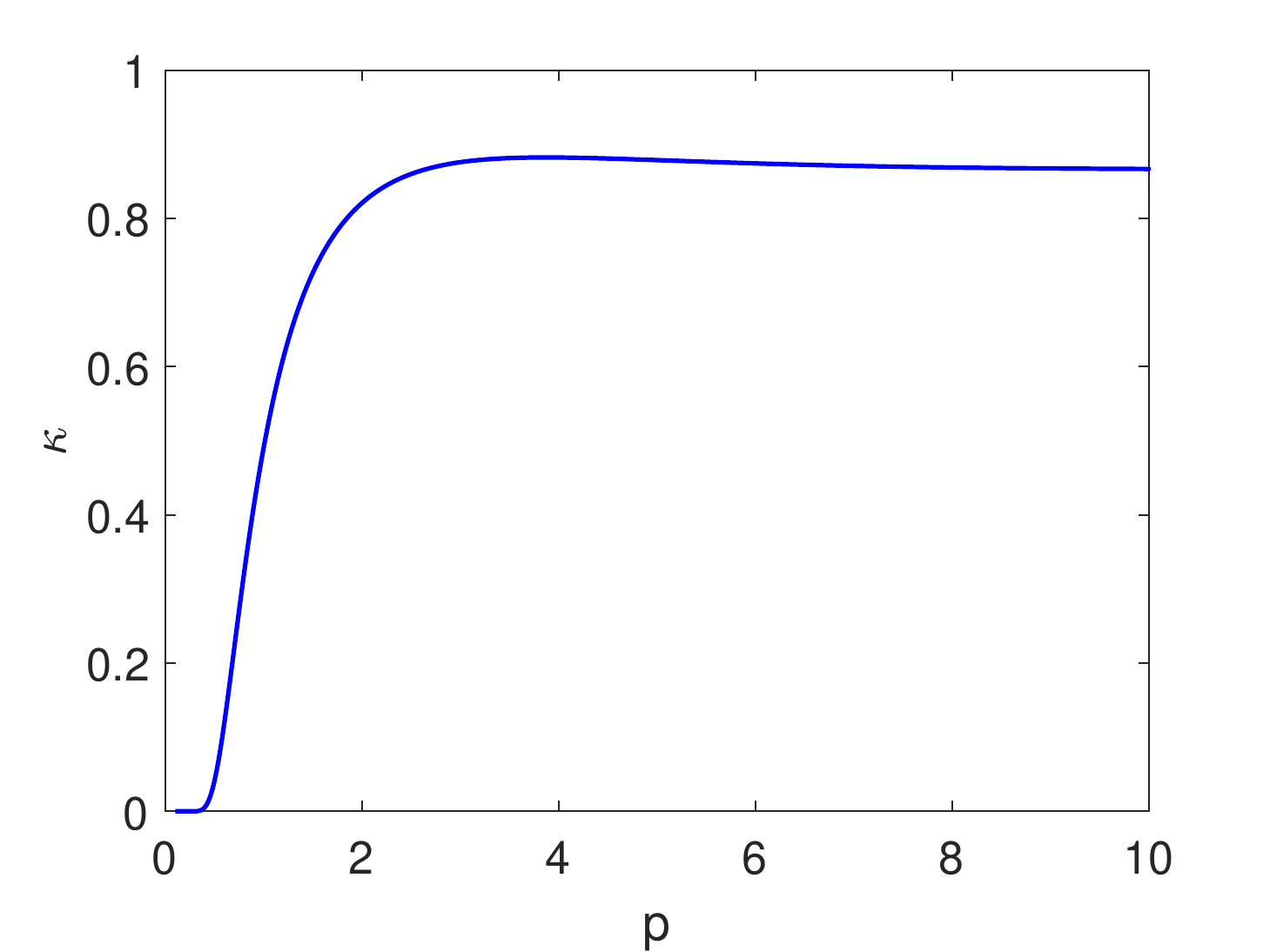}
	\includegraphics[width=0.450\textwidth]{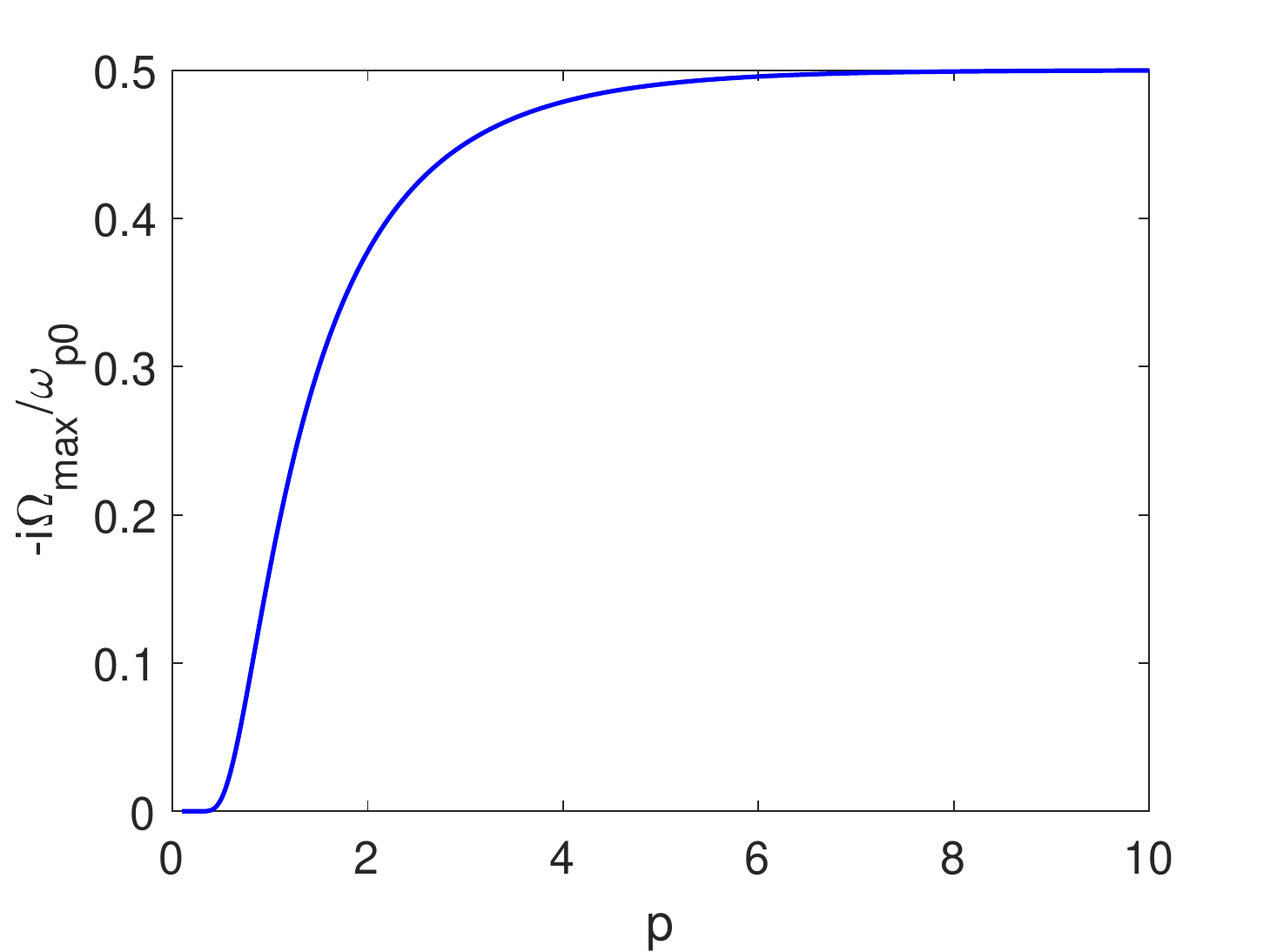}
	\caption{Wavenumber for the fastest growing mode (upper plot) and its growth rate (bottom plot) for different beam intensities.}
	\label{fig:2stream}
\end{figure}

The scaled wavenumber for the fastest growing mode $\kappa$ depends on a single parameter $p$, which describes the intensity of the beam. Note that the scaled intensity of the beam cannot be significantly smaller than unity for realistic beam and pipe radii as described by inequality (\ref{unstablecondition}). The plots for the wavenumber of the fastest growing mode and its growth rate are presented in Fig.~\ref{fig:2stream}. 
The results show a soft threshold for the instability at $p\approx 0.5$, which is not significantly different than the threshold described with Eq.~(\ref{unstablecondition}) due to presence of the pipe for realistic beam and pipe parameters in accelerators. The wavenumber for the fastest growing mode $\kappa\approx\sqrt{3/4}$ essentially matches the 1D result for intense enough beams, $p>2$. The growth rate of the instability also approaches 1D result $\Omega^2=-\omega_{p0}^2/4$ when the intensity of the beam increases.

In general case, the solution for the growth rate in two-stream instability is valid at large wavelengths, $p\kappa\ll1$ so that the electric field caused by the space charge is transversely uniform. However, the analysis in the short wavelength regime is valid if the growth length of the instability is much smaller than the beam beta function (characteristic length for beam defocusing).

\section{Summary}
\label{sec:conclusion}
We have developed a general formalism describing longitudinal space charge effects in beam-plasma systems. The self-consistent dynamics can be described in terms of effective longitudinal dielectric permittivity of the medium, which describes the response of the medium to the external space charge wave. The permittivity has a functional dependence matching the 1D plasma permittivity. The entire effect of the geometry (transverse density profile and presence of the conducting cylindrical wall) results in the effective plasma frequency being different from the 1D plasma frequency. The effective plasma frequency is described with Eq.~(\ref{omegapeff}).

The developed formalism provides a universal framework for studying longitudinal dynamics in various beam-plasma systems. Multiple species such as background plasma, beams with different energies, and different types of charged particles can be included in the analysis simultaneously as additive terms to the dielectric permittivity of the medium. Each of these species may have unique transverse profiles. Inclusion of kinetic effects is a straightforward generalization similar to the case of 1D plasma. A particle beam propagating through plasma affects the background density distribution and generates return current. These effects can also be included in the developed framework through introduction of additional particle species, which describe the return current and modified density profile.

\section{Acknowledgements}
Authors are thankful to Petr Anisimov, Stanislav Baturin, Trevor Burris-Mog, Dima Mozyrsky, Derek Neben, and Vitaly Pavlenko for fruitful discussions. Work supported by the US Department of Energy under contract number DE-AC52-06NA25396.



\begin{thebibliography}{8}
\bibitem{fast}
Fast space charge waves having $\omega/k>c$ will result in excitation of electromagnetic modes inside the vacuum pipe as they can be viewed as a source for Cherenkov radiation. Analysis of collective effects presented in this work will be incomplete in this case.
\end{thebibliography}
\end{document}